\documentclass[letterpaper,12pt]{article}   %% LaTeX 2e (preferred)
\usepackage{osajnl2} %% do not use with REVTeX4
\usepackage[draft]{hyperref} %% optional
\usepackage{graphicx}

\newcommand{\beq}{\begin{equation}}
\newcommand{\eeq}{\end{equation}}
\newcommand{\beqa}{\begin{eqnarray}}
\newcommand{\eeqa}{\end{eqnarray}}
\newcommand{\lam}{\lambda}

\newcommand{\rh}{\rho}

\newcommand{\si}{\sigma}

\def\pra#1{{ Phys.\ Rev. A\/} {\bf#1}}

\def\prl#1{{ Phys.\ Rev.\ Lett.} {\bf#1}}

\begin{document}

\title{Qubit Entanglement Driven by Remote Optical Fields}

%% For REVTeX it is possible to automate superscript and e-mail callouts with the superscriptaddress option; see REVTeX4 documentation.

\author{Muhammed Y\"ona\c{c},$^{1,*}$ and Joseph H. Eberly,$^2$}
\address{$^1$Department of Physics and Astronomy, University of
Rochester, New York 14627, USA }
\address{$^2$Rochester Theory
Center, and Department of Physics and Astronomy, University of
Rochester, New York 14627, USA }
\address{$^*$Corresponding author: yonac@pas.rochester.edu}

\begin{abstract}
We examine the entanglement between two qubits, supposed to be
remotely located and driven by independent quantized optical fields.
No interaction is allowed between the qubits, but their degree of
entanglement changes as a function of time.  We report a collapse
and revival of entanglement that is similar to the collapse and
revival of single-atom properties in cavity QED.
\end{abstract}


]

\begin{figure}[t]
\centerline{\includegraphics[width = 4 cm]{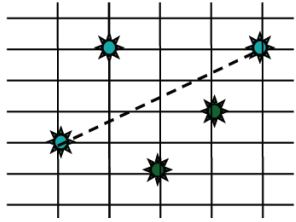}}
  \caption{\footnotesize  Sketch indicating non-interacting qubits in a
quantum storage network. Dashed line indicates two are
entangled.}\label{f.network-small}
\end{figure}

%=================================================================
%\section*{Introduction}
%================================================================
Control of the evolution of qubit entanglement and the time-dependent
behavior of qubit pairs in networks is relevant for quantum computing
and cryptography.  Entanglement will be stored in quantum memory
registers (see conceptual sketch in  Fig. \ref{f.network-small}) for
eventual use in some form of quantum communication. The qubits must
be controlled in some way externally, and we are interested here in
the response of the entanglement of a pair of stored qubits to
quantized optical control fields.

What we can call ``pure storage" requires mutual isolation and
non-interaction between qubits. Two-qubit evolution has been studied
without qubit isolation and usually allowing or relying on mutual
interactions to produce entanglement dynamics. Kim, et al., showed
\cite{knight} that an incoherent thermal field can create
entanglement between two such qubits.  Entanglement transfer between
two qubits and two separate fields was examined by Zhou and Wang
\cite{zhou}. In their treatment the quantum field was weak rather
than strong. The evolution of entanglement in a  qubit-field system,
where the qubit and the field start from mixed states was examined by
Rendell and Rajagopal \cite{rr}. They aimed to calculate the
entanglement embedded in the full system, and because of the lack of
an entanglement measure for $2\times\infty$ systems they calculated a
lower bound for the concurrence instead.

Other studies \cite{Yonac-etal06, Yonac-etal07} have shown that two isolated
qubits can exhibit periodic fluctuations in their entanglement in the
form of early-stage decoherence (ESD - also referred to as
entanglement sudden death) \cite{ye4} when the qubits are modelled as
``controlled" locally by interaction with only single photons.
However, manageable control fields are better modelled as containing
many photons. Here we retain a quantum picture of two many-photon
well-phased control fields by using a coherent state description of
them with a large mean photon number $\bar n \gg 1$.

%=================================================================
%\section*{Hamiltonian and Initial State}
%=================================================================

For our calculations we take single-mode control fields. Each field
is assumed, for simplicity, to be exactly resonant with the flip
transition between the ground $|g\rangle$ and excited $|e\rangle$
states of the qubit that it addresses. The well-known
Jaynes-Cummings (JC) interaction \cite{jc} is then relevant at each
qubit site (labelled i = 1,2), and the interaction Hamiltonian is
given by:

\begin{equation}
H_I=\sum_{i=1,2}\hbar g(a_i\sigma_i^{+} + a_i^{\dagger}\sigma_i^{-}),
\end{equation}
where $a_i$ and $a^{\dagger}_i$ are the photon annihilation and
creation operators for site $i$, and $\sigma_i^+$ and $\sigma_i^-$
are the raising and lowering Pauli matrices for atom $i$, and $g$ is
the coupling constant between atoms and fields, taken the same for
both for greatest simplicity hereafter.  For the two coherent state
fields we take the
same $\bar n$ for simplicity, and we assume initial entanglement in
the form of a familiar Bell State:
\beq \label{BellState}
|\Psi(0)\rangle = (|eg\rangle+|ge\rangle)/\sqrt{2}.
\eeq

%=================================================================
%\section*{Theory}
%=================================================================

Since JC time evolution is unitary, no information can be truly lost
in evolution via $H_I$, but the infinite range of photon numbers in a
coherent state brings aspects of open-system theory into play.
However, the quasi-classical nature of practically available control
fields suggests that we not expect their quantum characteristics to
be dominant. Thus we will trace out the fields and follow only the
qubit entanglements.

In this discussion we will use Wootters' concurrence \cite{wootters}
as our entanglement measure, which is given by
\beq \label{definitionc}
C(\rh) = \max\{0,\sqrt{\lam_1} - \sqrt{\lam_2} - \sqrt{\lam_3} -
\sqrt{\lam_4} \},
\eeq
where the quantities $\lam_i$ are the eigenvalues in decreasing order
of the matrix
\beq \zeta=\rho(\sigma_y\otimes
\sigma_y)\rho^*(\sigma_y\otimes \sigma_y).
\label{concurrence}
\eeq
Here $\rh$ is our two-qubit reduced-state density matrix, $\rh^*$
denotes the complex conjugation of $\rh$ in the standard basis, and
$\si_y$ is the Pauli matrix expressed in the same basis.

The photon number in a coherent field is Poisson distributed and
relatively tightly centered around $\bar{n}$ when $\bar n \gg 1$.
This suggests a shortcut approximation, to be checked numerically, in
which we represent the field density matrix as a Fock state having
photon number equal to $\bar n$. An important simplification
occurs in taking the same $\bar n$ for both control fields. Then the
initial field state is $|\bar n\rangle\otimes|\bar n\rangle$ and the
reduced density matrix for the qubits is:
\begin{equation}\label{Xstate}
\rho = \left(
\begin{array}{cccc}
      a & x & x & x \\
      x & b & z & x \\
      x & z* & c & x \\
      x & x & x & d \\
    \end{array}
\right) \to \left(
\begin{array}{cccc}
      a & 0 & 0 & 0 \\
      0 & b & z & 0 \\
      0 & z* & c & 0 \\
      0 & 0 & 0 & d \\
    \end{array} \right),
\end{equation}
where we have used the standard two-qubit  basis [$ee,\ eg,\ ge,\
gg$]. The elements indicated by $x$ are zero because of the
equal-$\bar n$ simplification.  Thus, under the assumptions
mentioned, $\rho$ is of $X$-type (see \cite{Yu-Eberly07}).

For an $X$-type $\rho$, Eq.(\ref{definitionc}) turns into:
\beq
C(\rho) = 2\ max[~ 0,\ |z|-\sqrt{ad}~ ].
\eeq
The control fields induce growth in time of the elements $a,\ d$,
which are the only ones not already present in the original maximally
entangled state (\ref{BellState}). Their growth and any decline of
$z$ will cause entanglement to decrease.

Having used the Fock state shortcut to obtain (\ref{Xstate}), we
avoid using it further now and calculate the elements $z,\ a,\ d$ for
the coherent state. We introduce the Poisson number distribution by
the coherent-state amplitude measure $A_n =
e^{-|\alpha|^2/2}\alpha^{n}/\sqrt{n!}$, where $|\alpha|^2 = \bar n$,
we obtain a doubly infinite series summation:
\beqa
z &=& \frac{1}{2}\Big\{\sum_{n,m}A_n^2A_m^2C_nC_{n+1}C_mC_{m+1}\nonumber\\
&-&A_nA_{n-1}A_mA_{m+1}S_nC_{n+1}C_mS_{m+1}\nonumber\\
&+&A_nA_{n-2}A_mA_{m+2}S_nS_{n-1}S_{m+1}S_{m+2}\nonumber\\
&-&A_nA_{n-1}A_mA_{m+1}S_nC_{n-1}S_{m+1}C_{m+2}\Big\},
\eeqa
where $C_n=\cos(gt\sqrt{n})$ and $S_n=\sin(gt\sqrt{n})$.

Similarly the series summations for $a$ and $d$ are;
\beqa
a &=& \frac{1}{2}\Big\{\sum_{n,m}A_n^2A_m^2C_{n+1}^2S_m^2\nonumber\\
&+&A_nA_{n+1}A_mA_{m-1}S_{n+1}C_{n+1}S_mC_m\nonumber\\
&+&A_n^2A_m^2S_n^2C_{m+1}^2\nonumber\\
&+&A_nA_{n-1}A_mA_{m+1}S_nC_nS_{m+1}C_{m+1}\Big\}
\eeqa
and
\beqa
d &=& \frac{1}{2}\Big\{\sum_{n,m}A_n^2A_m^2S_{n+1}^2C_m^2\nonumber\\
&+&A_nA_{n+1}A_mA_{m-1}S_{n+1}C_{n+1}S_mC_m\nonumber\\
&+&A_n^2A_m^2C_n^2S_{m+1}^2\nonumber\\
&+&A_nA_{n-1}A_mA_{m+1}S_nC_nS_{m+1}C_{m+1}\Big\}. \eeqa

The infinite extent of these summations of course reflects the fact
that we have coupled the qubits to an infinite state space. The sums
cannot be analytically completed, but in our calculations we choose
$\alpha = 10$, i.e., $\bar n = 100 \gg 1$, so we can obtain good
approximations if we use Stirling's formula for $n!$,
\beq
n!=\sqrt{2\pi n} n^n e^{-n},
\eeq
and Euler's formula to approximate the terms in the summations above
by integrals.  If we approximate the terms like
$A_nA_{n+1}A_mA_{m-1}$ with $A_n^2 A_m^2$, which introduces an error
of order $1/\bar{n}$ near the Poisson peaks $n \approx m \approx
\bar n$, helpful cancellations can be identified, and we obtain an
approximate expression for $|z|-\sqrt{ad}$:
\beqa \label{eq.Lambda}
|z|-\sqrt{ad} &\cong& \frac{1}{4}\Big[e^{-g^2t^2/8\bar{n}^2} - 1
\Big]\nonumber\\
& + &\frac{1}{2}\Big[\sum_{n}A_n^2 \cos(2gt\sqrt{n})\Big]^2 \nonumber\\
& - & \frac{1}{2}\Big[\sum_n A_n^2 \sin(2gt\sqrt{n})\Big]^2.
\eeqa

\begin{figure}[h]\begin{center}
    \includegraphics[width=8.0cm]{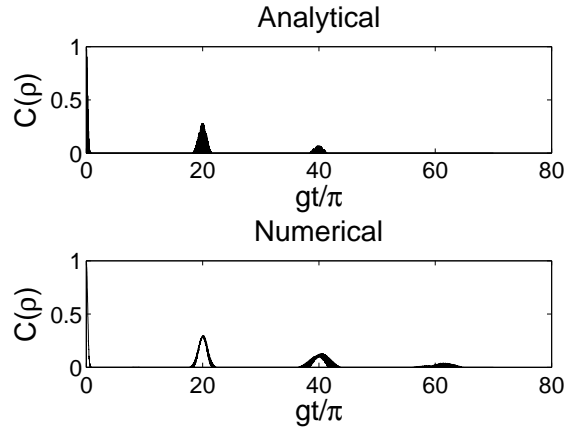}\\
    \end{center}
    \caption{\footnotesize The analytical and numerical results for
entanglement. As expected, revivals occur, which are predicted by the
approximate analytical results reasonably well. Better resolution of
the rapid oscillations is provided in Fig. \ref{f.num_analyt_detail}.
}\label{f.num_analyt}
\end{figure}

The summations in (\ref{eq.Lambda}) involving $\cos(2gt\sqrt{n})$
and $\sin(2gt\sqrt{n})$ also cannot be completed analytically, but
are the same type as those for qubit inversion in the original
discussion of quantum revivals \cite{eb81} for zero detuning, so we
expect to see similar revival behavior here. Fig. \ref{f.num_analyt}
shows our analytical results for the function
$2(|z|-\sqrt{ad})$ in comparison with the complete numerical
evaluation of Eq.(\ref{definitionc}) without making the $X$-type approximation.

We note that both curves in Fig. \ref{f.num_analyt} indicate repeated
occurrences of early-stage decoherence (ESD). The recurring positive
entanglement events are not periodic in amplitude, as found in the
weak-field cases discussed in  \cite{Yonac-etal06, Yonac-etal07}, but
are periodic in time. Fig. \ref{f.num_analyt_detail} below shows that
the analytic approximation includes micro-ESD events that are not
present in the full expression.

Because of the exponential term in (\ref{eq.Lambda})  the envelope of
the function decreases slowly. Also, close to the revival regions the
function makes rapid oscillations with period $\tau =
\pi/(2g\sqrt{\bar{n}})$ which is half the corresponding period in the
inversion case. The reason is that in (\ref{eq.Lambda}) we have
squares of summations rather than the summations for the original
inversion calculations \cite{eb81}.  As the figure shows, our
analytic summations and the approximations they are based on work
rather well, capturing all major aspects of the full numerical
result. The function $2(|z|-\sqrt{ad})$ can suitably predict the
magnitude and revival of the entanglement while not a perfect
substitute for it.

\begin{figure}[h]\begin{center}
    \includegraphics[width=8.0cm]{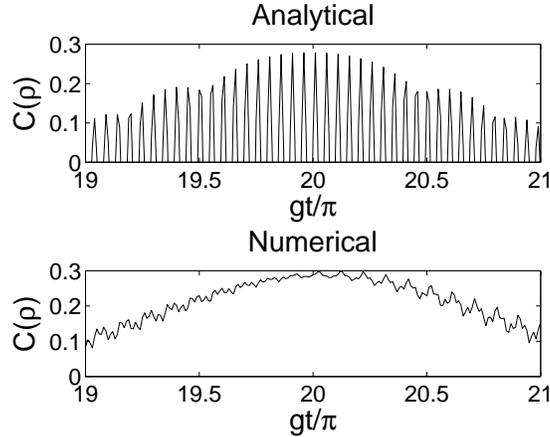}\\
    \end{center}
    \caption{\footnotesize A more detailed plot of the results around
$t=20\pi/g$
shown in Fig. \ref{f.num_analyt}.  Analytical results are for the
$X$-type $\rho$ while the numerical ones are for the original $\rho$.
} \label{f.num_analyt_detail}
\end{figure}

%=================================================================
%\section*{Conclusion-new}
%=================================================================

In conclusion, we have made a numerical and analytic examination of
the entanglement dynamics between two qubits controlled by (or,
receiving ``instructions" from) quantized optical fields that are
modelled as equal-amplitude coherent state modes.  We show that
periodic ESD effects, as were predicted for very weak control fields
in \cite{Yonac-etal06, Yonac-etal07}, are still present but in
substantially modified form. Many very rapid appearances of the ESD
effect are contained in the top plot of Fig.
\ref{f.num_analyt_detail} and are not present in the numerical
evaluation of concurrence for the complete two-qubit reduced $\rho$.
Both the X-state simplification in format and the high-$\bar n$
approximation contribute to the differences, but the latter is much
less significant when $\bar n \gg 1$, as here. A visible consequence
in both plots coming just from the open-system nature of the coherent
states is the imperfect recovery of entanglement in each successive
revival zone.

Figs. \ref{f.num_analyt} and \ref{f.num_analyt_detail} have
implications for further work. They show that the major features of
the entangled qubits' response are very well captured even by the
rather severe X-state simplification we introduced. The benefits of
having analytic expressions, as provided by the X-state format, can
be expected to be substantial in guiding and interpreting more
complex calculations.

In this regard, we believe it will be interesting to investigate
dynamic entanglement behavior under different assumptions about the
state of the system.  For example, the response to squeezed fields or
fields acting at different times or fields of significantly different
intensity or mode frequency are open for study. Similarly, the state
reduction employed to reach the X state is not limited to two-qubit
situations. These expanded topics will be the focus of a wider
investigation.

Acknowledgement: This work was supported by ARO Grant
W911NF-05-1-0543.

\end{document}